\documentclass[showpacs,amsmath,amssymb,aps,prl]{revtex4}

\usepackage{graphicx}
\usepackage{dcolumn}
\usepackage{bm}

\begin{document}

\preprint{}

\title{Aggregation kinetics in a model colloidal suspension}

\author{Sorin Bastea}
\email{sbastea@llnl.gov}
\affiliation{Lawrence Livermore National Laboratory, P.O. BOX 808, Livermore, CA 94550}


\begin{abstract}
We present molecular dynamics simulations of aggregation kinetics in a 
colloidal suspension modeled as a highly asymmetric binary mixture. 
Starting from a configuration with largely uncorrelated colloidal particles 
the system relaxes by coagulation-fragmentation dynamics to a structured 
state of low-dimensionality clusters with an exponential size distribution. 
The results show that short-range repulsive interactions alone can give 
rise to so-called cluster phases.
For the present model and probably other, more common colloids, 
the observed clusters appear to be equilibrium 
phase fluctuations induced by the entropic inter-colloidal attractions.
\end{abstract}


\maketitle

Colloidal suspensions are common in biological 
and geological systems and their unusual properties are intensely 
exploited in many industrial applications, from pharmaceutics to food 
processing. Although some, e.g. noble metal colloids, have been produced and 
used for centuries, the systematic study of these systems is a more recent 
endeavor \cite{al02}. One of the remarkable features of colloids is the possibility 
of tuning the colloidal interactions \cite{yb03}. This multiplies 
the already rich physical behavior exhibited by colloids  
\cite{ssc93,wn03,hg03,af03,ssc04,ll04,mdh04,pfc04,mct04,sdh04,fw05} and significantly 
increases their technological potential, but it also underscores the need 
for understanding the general principles and mechanisms that govern their 
equilibrium and kinetic behavior. The low volume fractions 
kinetic arrest leading to gelation in weakly attractive hard-sphere colloids 
has been found for example to share many of the features of the high volume 
fractions colloidal glass transition \cite{sps01}. The aggregation of colloids 
and the ensuing highly structured states such as gels are a direct consequence 
of the short-ranged attractions typically encountered in these systems, known 
as depletion or entropic interactions \cite{ao54}. The role of repulsions 
however is also important, and recently much attention has been 
devoted to elucidating the effect of long-range repulsions in the 
formation of cluster phases and the gelation transition \cite{ssc04,smz04,cad05}. 
Understanding clustering effects and the mechanisms that control 
them is important conceptually as well as practically in systems ranging from 
biological to engineered ones, e.g. nanofluids.
In the present paper we explore using molecular dynamics (MD) simulations 
the characteristics of aggregation kinetics in a model colloidal suspension 
with short-range repulsions that explicitly includes both colloidal and solvent particles. 
We find that the system evolution is well described as coagulation-fragmentation 
dynamics surprisingly similar to a simple polymerization-depolymerization 
process. The final state is a structured phase of low-dimensionality 
clusters with an exponential size distribution. These properties appear to be all 
due to the bond-like character of the induced short-ranged inter-colloidal 
attractions.

The simplest instance of a colloidal suspension is perhaps a binary hard-sphere 
mixture with significant size asymmetry. Dijkstra et al. \cite{dre99} have shown that 
the phase behavior of such a system can be accurately predicted by integrating out 
the effect of the small particles and introducing an effective depletion potential 
\cite{ao54} between the large (colloidal) particles. This approach provides 
significant insight into the structure of the phase diagram and simplifies considerably 
the computational study of the mixture. However, it may not be necessarily 
appropriate for the study of non-equilibrium processes in colloids. 
The motion of colloidal particles suspended in viscous liquids is generally coupled, 
even at large separations, 
by the induced suspending-liquid flows. These so-called hydrodynamic interactions 
are believed for example to play an important role in the aggregation kinetics \cite{ta00}. 
In the following we consider and fully model using MD simulations a system 
related to the binary hard-sphere mixture, 
consisting of two types of particles, {\it 1} - colloid
and {\it 2} - solvent, with equal masses $m$. The interactions between the 
particles are based on the inverse-12, 'soft-sphere' potential, whose 
properties have been well studied  \cite{hrj70}
\begin{equation}
u(r)=\epsilon\left(\frac{\sigma}{r}\right)^{12} 
\end{equation}
, and which we truncate and shift at $r/\sigma=2$. (We also define $u(r)=\infty$ for $r<0$.) 
The interactions are:
\begin{subequations}
\begin{eqnarray}
&&u_{11}(r) = u(r - 2R_c)\\
&&u_{12}(r) = u(r - R_c)\\
&&u_{22}(r) = u(r)
\end{eqnarray}
\label{eq:ur}
\end{subequations}
Similar potentials, that take into account the 'size' of the colloidal particles by 
introducing a hard core radius $R_c$, have been employed before to model suspensions 
\cite{nmh98}. For temperatures $k_B T\simeq \epsilon$ the effective diameters 
corresponding to the above interactions should be well approximated by $\sigma_2=\sigma$, 
$\sigma_{12}=R_c+\sigma$, and $\sigma_1\equiv\sigma_c=2R_c + \sigma$, and satisfy 
additivity, $\sigma_{12} = (\sigma_1 + \sigma_2)/2$.
The particular system that we focus on has a colloid-solvent 'diameter' ratio $\sigma_c/\sigma=5$, 
colloid 'volume fraction' $\phi_c=\pi n_c{\sigma_c}^3/6=0.1$ and 
solvent 'volume fraction' $\phi=\pi n{\sigma^3}/6=0.37$; $n_c$ and $n$ 
are the corresponding number densities, $n_c=N_c/V$, $n=N/V$. We perform MD 
simulations of this system in the microcanonical (NVE) ensemble with $N_c=200$, $N\simeq 10^5$ and an 
average temperature set to $k_B T=\epsilon$. Such simulations of binary mixtures 
with large size ratios have been done before \cite{jrs87}, but not, to our knowledge, 
for colloid volume fractions as small as the present one.

The simulations are initiated from a configuration of non-overlapping 
(as defined by the effective diameters), but otherwise randomly distributed 
colloid and solvent particles, and the system is equilibrated at $k_BT=\epsilon$ 
using a combination of velocity rescaling and Anderson thermostatting. 
Thermal equilibrium is achieved over time scales much shorter than the 
scales over which the colloidal particles move over any 
significant distances, about $5000$ MD steps, leading 
to a state corresponding to a fully equilibrated solvent and essentially 
uncorrelated colloidal particles. A largely similar 
configuration would be expected for example in the case of a hard-sphere colloid 
shortly after a rapid quench from a high temperature state. 
The system evolution is subsequently followed using conventional (constant energy) MD for 
over $5\times 10^5$ time steps.

The evolving configuration of the large (colloidal) particles during the simulations 
is best characterized by using a standard aggregation criterion, that places two 
particles in the same cluster when their relative distance is smaller than a certain value 
$r_b$. For the present work we set $r_b=1.05\sigma_c$, but our conclusions are independent 
of this value in a 
fairly wide range. The application of this criterion yields a time dependent cluster 
size distribution $N_k(t)$, or in scaled form $n_k(t)=N_k(t)/N_c$, where 
$N_c=\sum_k k N_k(t)$ is the (constant) total number of particles. 
We show in Fig. 1 the evolution of the (normalized) number of monomers 
$n_1(t)$, along with the inverse of the (normalized) total number 
of clusters, $s_1(t)=\sum_k k N_k(t)/\sum_k N_k(t)$; $s_1$ can also be interpreted 
as a measure of the average cluster size. The number of monomers in the system 
steadily decreases, while the cluster size increases, with both quantities 
apparently reaching 
at late times well defined plateaus. Visual inspection of the 
clustering dynamics reveals tenuous structures reminiscent of branched polymers - 
see Fig. 2 - whose evolution 
is driven by aggregation and breakup processes. These features suggest that general 
concepts developed to describe coagulation-fragmentation kinetics \cite{fmd86,szt87} 
may also be relevant to the present system. 
The starting point for this analysis 
is the generalized Smoluchowski equation for the cluster size distribution:
\begin{eqnarray}
\frac{dn_k}{dt}=\frac{1}{2}\sum_{i+j=k}[K_{ij}n_in_j-F_{ij}n_{i+j}]
-\sum_{j=1}^{\infty}[K_{kj}n_kn_j-F_{kj}n_{k+j}]
\end{eqnarray}
with coagulation and breakup rate coefficients $\{K_{ij}\}$ and $\{F_{ij}\}$, $i,j=1, \infty$. 
The cluster population $n_k(t)$ whose evolution is described by the above 
equations is generally assumed to satisfy a dynamical scaling 
relation \cite{fmd86,szt87}
\begin{equation}
n_k(t)=s(t)^{-2}\phi(x)
\label{eq:ds}
\end{equation}  
where $x=k/s(t)$, $s(t)$ is a characteristic cluster size 
and $\phi(x)$ a scaling function dependent on the details of process, 
i.e. the coefficients $\{K_{ij}\}$ and $\{F_{ij}\}$. 
The applicability of Eq. \ref{eq:ds} to systems that reach equilibrium, i.e. 
where $s(t)$ goes to a constant value when $t\rightarrow\infty$, appears to 
be somewhat more 
limited \cite{vz88} than initially assumed. Nevertheless, under certain reasonable 
assumptions for the coagulation and breakup rates 
and in particular for a classical model of polymerization-depolymerization with 
constant coefficients \cite{bt45} dynamical scaling holds exactly. 
For small deviations from equilibrium it can also be shown \cite{szt87} that 
dynamical scaling implies that $s(t)$, and therefore $s_1(t)$, relaxes exponentially 
to its final, equilibrium value \cite{szt87}. Furthermore, for the case of 
Ref. \cite{bt45} the product $n_1(t)s_1^2(t)$ is time independent, 
$n_1(t)s_1^2(t)=1$, while in general should increase only slowly in a narrow range. 

For the asymmetric binary mixture considered here as an archetype of a colloidal
suspension we find that $s_1(t)$ relaxes to a very good approximation in an 
exponential fashion, and
moreover that the relation $n_1(t)s_1^2(t)=1$ is very close to being satisfied - see Fig. 1.
This makes the behavior of this simple model, containing only short-range repulsive 
potentials, surprisingly similar to that of common associating systems. 
As a test of dynamical scaling we consider the usual
measure of the mean cluster size \cite{fmd86,szt87}, $s(t)=\sum_k k^2 n_k(t)$, which
depends linearly on $s_1(t)$ for the model of Blatz and Tobolsky \cite{bt45}, and in general
should be proportional with it when $s$ is large enough. This is also rather well satisfied,
see Fig. 1(inset). 

The morphology of the typical clusters - see Fig. 2 - closely resembles the open
structures usually encountered in colloidal aggregation \cite{wo84,ta00}. Although
the observed assemblies are not nearly as large as those obtained upon irreversible
flocculation \cite{wo84}, their dimensionality can be well quantified
	       by considering their radius of gyration
$kR^2_g(k)=\langle\sum^k_{i=1} [(x_i - x_{CM})^2 + (y_i - y_{CM})^2 + (z_i - z_{CM})^2]\rangle$,
where $i$ denotes the particles in a cluster of size $k$, $CM$ is the cluster center of mass, 
and the average is taken over the final population of clusters of size $k$. The resulting radii
- see Fig. 3 - obey $R_g(k)\propto k^{1/D_f}$, with $D_f\simeq 2$. The fractal dimension $D_f$
is slightly higher than the one observed for typical colloidal aggregates 
\cite{wo84}, but in agreement with that of randomly branched polymers \cite{ps81}.

The cluster size distribution in the final state - $n_{eq}(k)$, see Fig. 4 - compares 
well with the distributions obtained in model coagulation-fragmentation 
dynamics \cite{fmd86,bt45} and may be reasonably described at large cluster sizes 
as a decaying exponential. From the point of view of the kinetics this appears to be a consequence 
of an approximately cluster-independent break-up rate \cite{fmd86,bt45}, 
which is consistent with the low cluster dimensionality and the equivalence for this 
system of the 'bonds' that hold together structures of different sizes.
In fact, thermodynamic arguments applicable to equilibrium 
polymers systems \cite{cc90} yield a similar prediction, while general concepts 
pertaining to phase fluctuations 
(and nucleation) in solutions \cite{ll}, would also predict an exponentially decreasing abundance 
of large clusters for single-phase equilibrium close to a two-phase coexistence region. 
Nevertheless, since the size of our system in 
terms of the number of large particles modeled is fairly small, we further address this issue 
by considering an equivalent one-component model with effective interactions.
 
To this end we consider the structure of the mixture, in particular the colloid-colloid 
pair correlation function, $g_{11}(r)$ - Fig. 5. This function 
exhibits the sharp nearest-neighbor peak also encountered in the effective one-component 
modeling of asymmetric binary hard-sphere mixtures \cite{dre99}, which is due to 
the strong, entropically induced attractions. However, $g_{11}(r)$ is also considerably 
more structured than such one-component correlation functions due 
to the explicit presence of the solvent. For our own single component modeling 
we consider particles interacting through a potential $V_{eff}(r)$ defined by 
$\beta V_{eff}(r)=-\ln[g_{11}(r)]$, i.e. the potential of mean force, but which is truncated 
after its first maximum 
at a distance $r_m$ satisfying $V_{eff}(r_m)=0$. We expect the resulting 
interaction to be an appropriate effective pair potential for the rather low colloid volume 
fraction system studied here. $V_{eff}$ has a fairly deep well followed by a 
barrier - see Fig. 5 (inset) - and is analogous with the effective potentials 
used to study the phase diagram of asymmetric binary hard-sphere mixtures \cite{dre99}.
MD simulations with $864$ particles interacting through this potential at a 
temperature $k_BT=\epsilon$ were carried out. The system 
requires very long equilibration times, of the order of $10^6$ steps, which we then follow by 
standard (NVE) MD to determine the fluid structure. The main features of the resulting 
pair correlation 
function - Fig. 5 - are not surprising. The sharp nearest neighbor peak of the 
binary mixture along with the next (colloid-separated) nearest neighbor 
maximum are very well reproduced, but the intermediate structure due to the small 
particles is missing. To further elucidate the one-component system structure and compare it 
with that of the binary mixture we perform the same clustering 
analysis as before, using the same distance $r_b$. This yields the cluster 
population shown in Fig. 4, which agrees well with 
the fully modeled mixture result and exhibits a rather clear exponential decay at 
large cluster sizes. Somewhat surprisingly, the dimensionality of the 
clusters so obtained is also in very good agreement with that of the clusters encountered 
in the two-component simulation of the mixture. This appears to cast some doubt 
on the role played by hydrodynamic interactions in the formation of open 
structures \cite{ta00}.

The present MD results and analysis suggest that in strongly asymmetric binary mixtures 
exclusion-like interactions alone can give rise to highly structured, so-called cluster phases. 
For the binary mixture studied here as a paradigm of a colloidal suspension the clustering 
kinetics is dominated by two-body coagulation and fragmentation processes. This may be 
expected to hold with increasing accuracy as the colloidal volume fraction decreases. 
The resulting colloidal clusters are tenuous, open structures that can be classified as 
randomly branched polymers, which is largely a consequence of the very short range 
of the induced inter-colloidal attractions and not hydrodynamic interactions. In the final 
state the clusters are exponentially distributed in size for both the fully modeled mixture 
and an equivalent effective one-component fluid. This is consistent with single-phase 
equilibrium of the mixture close to a two-phase region of its phase diagram. The observed 
cluster 'phase' is then simply a signature of equilibrium phase fluctuations, controllable 
for example by temperature and pressure changes. 
This may also be the case with other, more standard examples of colloids where clusters 
form through reversible aggregation and are quite small in size \cite{ssc04}. When 
considered together with the observed properties of the aggregation kinetics these 
features lend support to the idea \cite{szt87} that the relaxation dynamics in such 
systems is well described as coagulation-fragmentation kinetics of phase fluctuations. 

This work was performed under the auspices of the U. S. Department of Energy by 
University of California Lawrence Livermore National Laboratory under Contract 
No. W-7405-Eng-48.

\begin{figure}
\includegraphics{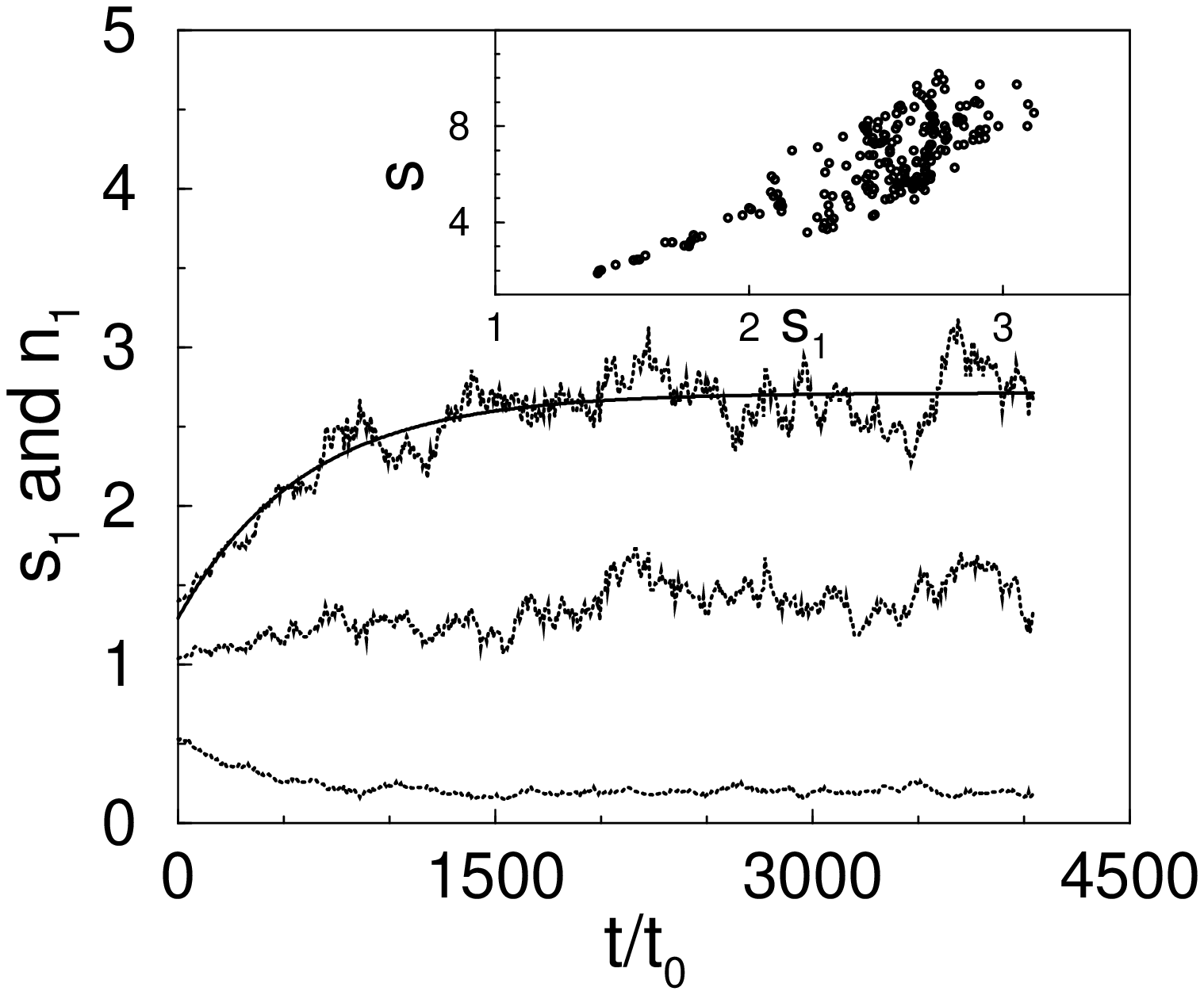}
\caption{Time evolution of the (normalized) number of monomers - $n_1$ (bottom dotted line), 
inverse of the (normalized) total number of clusters - $s_1$ (top dotted line), and $n_1 s_1^2$ 
(middle dotted line); solid line is an exponential relaxation fit; 
$t_0=\sigma(m/\epsilon)^{1/2}$. Inset: characteristic cluster size $s$ 
(see text) versus $s_1$.}
\label{fig:fig1}
\end{figure}

\begin{figure}
\includegraphics{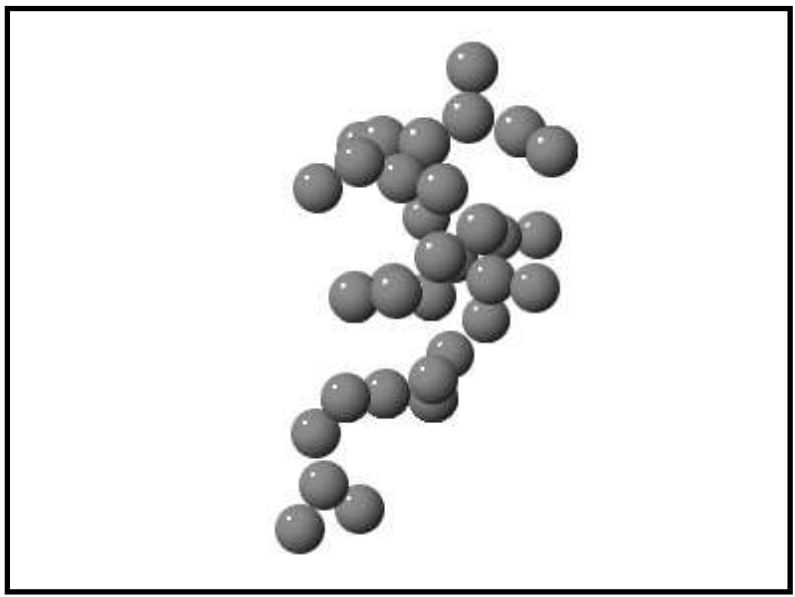}
\caption{A cluster of 33 colloidal particles.}
\label{fig:fig3}
\end{figure}

\begin{figure}
\includegraphics{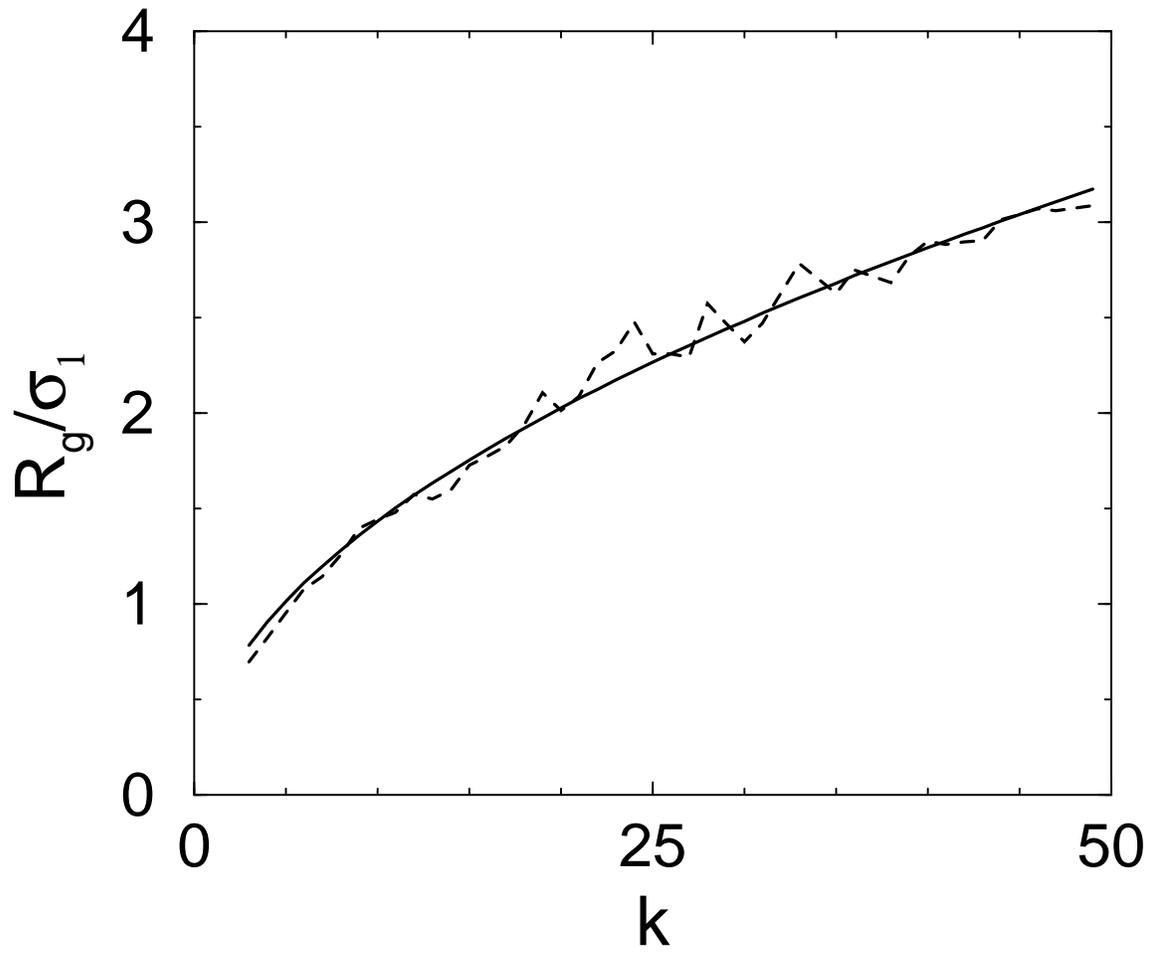}
\caption{Radius of gyration $R_g$ (see text) as a function of cluster size $k$: from 
simulations (dashed line) and power law ($R_g\propto k^{1/D_f}$) fit (solid line); $D_f=2$.}
\label{fig:fig2}
\end{figure}

\begin{figure}
\includegraphics{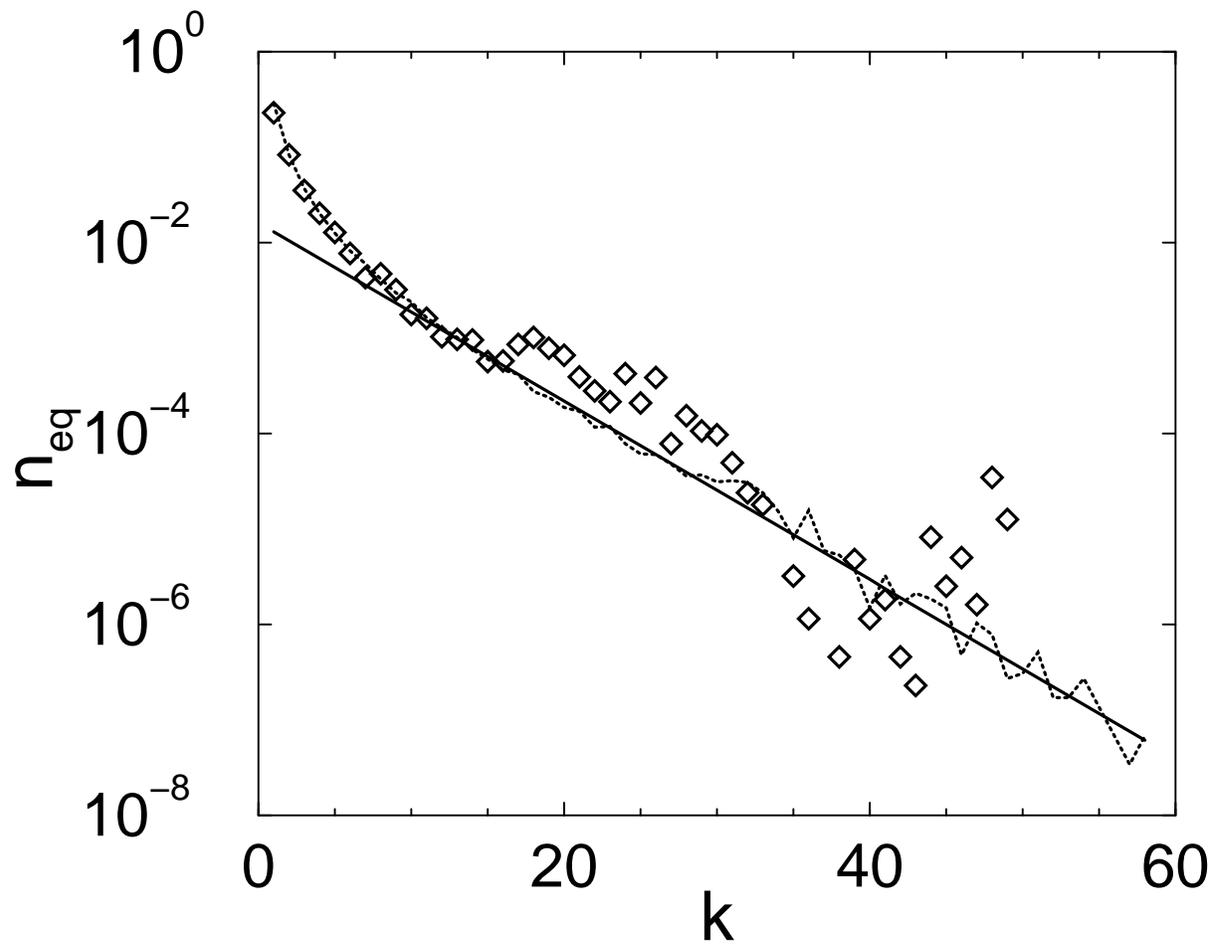}
\caption{Cluster size distribution from MD simulations of the 
binary mixture (diamonds), simulations of the one-component fluid (dotted line) 
and exponential decay fit (solid line).}
\label{fig:fig4}
\end{figure}

\begin{figure}
\includegraphics{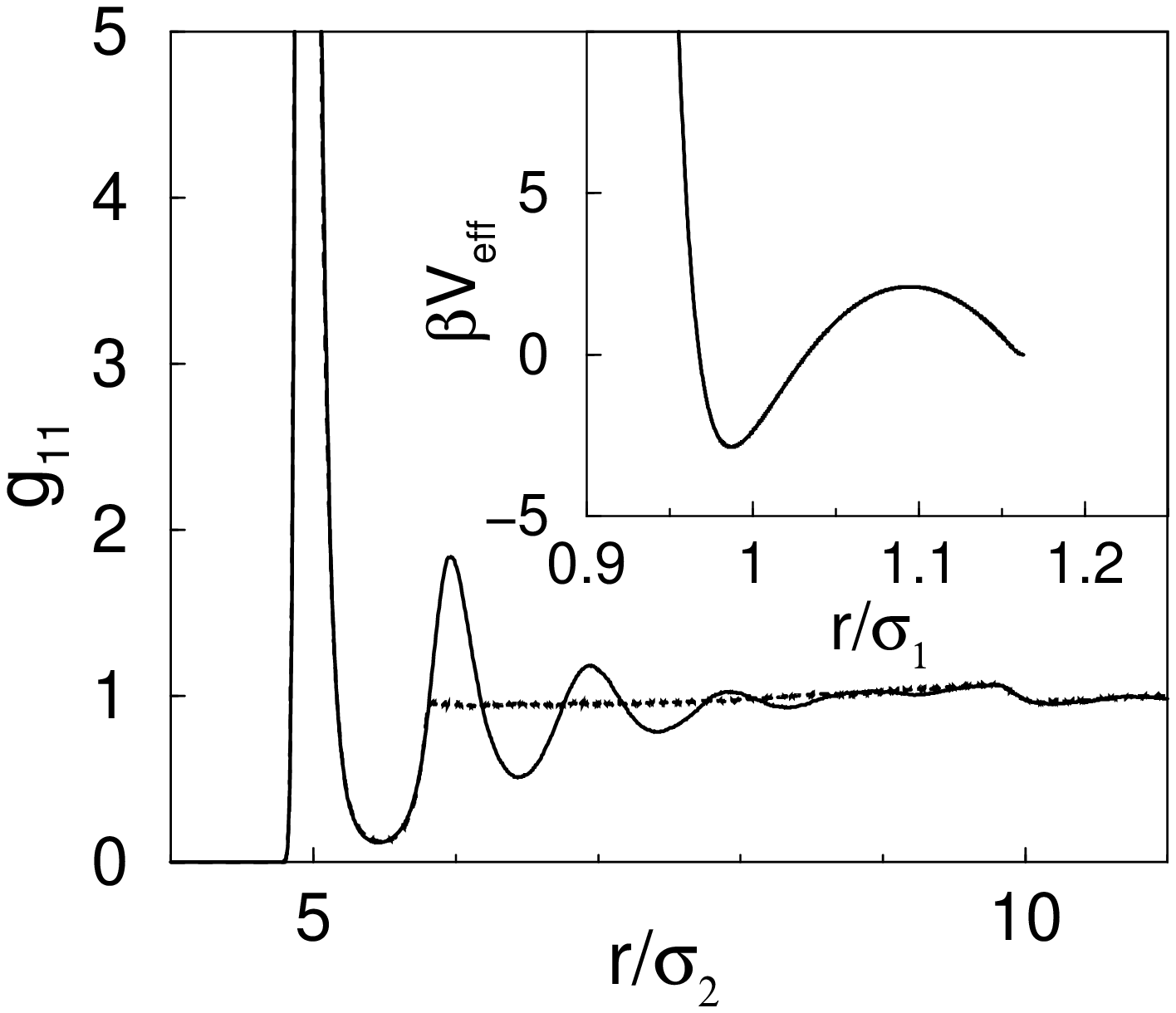}
\caption{Colloid-colloid pair correlation function of the binary mixture (solid line) 
and effective one-component system (dotted line). Inset: interaction potential $V_{eff}$ 
(see text) of the one-component system; $\beta=1/k_BT$.}
\label{fig:fig5}
\end{figure}


\begin{thebibliography}{99}
\bibitem{al02}For a recent review see, e.g., V.J. Anderson, H.N.W. Lekkerkerker, 
Nature {\bf 416}, 811 (2002).
\bibitem{yb03}A. Yethiraj, A. van Blaaderen, Nature {\bf 421}, 513 (2003).
\bibitem{ssc93}I.M. de Schepper, H.E. Smorenburg, E.G.D. Cohen, Phys. Rev. Lett. {\bf 70}, 
2178 (1993).
\bibitem{wn03}D.T. Wasan, A.D. Nikolov, Nature {\bf 423}, 156 (2003).
\bibitem{hg03}Y. Han, D.G. Grier, Phys. Rev. Lett. {\bf 91}, 038302 (2003).
\bibitem{af03}S. Auer, D. Frenkel, Phys. Rev. Lett. {\bf 91}, 015703 (2003).
\bibitem{ssc04}A. Stradner, H. Sedgwick, F. Cardinaux, W.C.K. Poon, S.U. Egelhaaf, 
P. Schurtenberger, Nature {\bf 432}, 492 (2004).
\bibitem{ll04}J. Liu, E. Luijten, Phys. Rev. Lett. {\bf 93}, 247802 (2004).
\bibitem{mdh04}M.D. Haw, Phys. Rev. Lett. {\bf 92}, 185506 (2004).
\bibitem{pfc04}A.M.Puertas, M. Fuchs, M.E. Cates, J. Chem. Phys. {\bf 121}, 2813 (2004).
\bibitem{mct04}S. Manley et al., Phys. Rev. Lett. {\bf 93}, 108302 (2004).
\bibitem{sdh04}M. Schmidt, M. Dijkstra, J.-P. Hansen, Phys. Rev. Lett. {\bf 93}, 088303 (2004).
\bibitem{fw05}J. Forsman, C.E. Woodward, Phys. Rev. Lett. {\bf 94}, 118301 (2005).
\bibitem{sps01}P.N. Segr\'{e}, V. Prasad, A.B. Schofield, D.A. Weitz, Phys. Rev. Lett. {\bf 86}, 
6042 (2001).
\bibitem{ao54}S. Asakura, F. Oosawa, J. Chem. Phys. {\bf 22}, 1255 (1954).
\bibitem{smz04}F. Sciortino, S. Mossa, E. Zaccarelli, P. Tartaglia, Phys. Rev. Lett. {\bf 93}, 
055701 (2004).
\bibitem{cad05}A.I. Campbell, V.J. Anderson, J.S. van Duijneveldt, P. Bartlett, 
Phys. Rev. Lett. {\bf 94}, 208301 (2005).
\bibitem{dre99}M. Dijkstra, R. van Roij, R. Evans, Phys. Rev. E {\bf 59}, 5744 (1999).
\bibitem{ta00}H. Tanaka, T. Araki, Phys. Rev. Lett. {\bf 85}, 1338 (2000). 
\bibitem{hrj70}W.G. Hoover, M. Ross, K.W. Johnson, D. Henderson, J.A. Barker, 
B.C. Brown, J. Chem. Phys. {\bf 52}, 4931 (1970). 
\bibitem{nmh98}M.J. Nuevo, J.J. Morales, D.M. Heyes, Phys. Rev. E {\bf 58}, 5845 (1998).
\bibitem{jrs87}G. Jackson, J.S. Rowlinson, F. van Sol, J. Phys. Chem. {\bf 91}, 4907 (1987).
\bibitem{fmd86}F. Family, P. Meakin, J.M. Deutch, Phys. Rev. Lett. {\bf 57}, 727 (1986).
\bibitem{szt87}C.M. Sorensen, H.X. Zhang, T.W. Taylor, Phys. Rev. Lett. {\bf59}, 363 (1987).
\bibitem{vz88}R.D. Vigil, R.M. Ziff, Phys. Rev. Lett. {\bf 61}, 1431 (1988).
\bibitem{bt45}P.J. Blatz, A.V. Tobolsky, J. Phys. Chem. {\bf 49}, 77 (1945).
\bibitem{wo84}D.A. Weitz, M. Oliveria, Phys. Rev. Lett. {\bf 52}, 1433 (1984).
\bibitem{ps81}G. Parisi, N. Sourlas, Phys. Rev. Lett. {\bf 46}, 871 (1981).
\bibitem{cc90}M.E. Cates, S.J. Candau, J. Phys.: Condens. Matter {\bf 2}, 6869 (1990).
\bibitem{ll}See, e.g., L.D. Landau, E.M. Lifshitz, {\it Statistical Physics}, $3^{rd}$ edition,
(Butterworth-Heinemann, Oxford, 1980).

\end{thebibliography}
\end{document}